\documentclass[prc,aps,nofootinbib,showkeys,showpacs,twocolumn]{revtex4}
\usepackage{epsfig}
\usepackage{graphicx}
\usepackage{amssymb}
\usepackage{color}
\begin{document}

\title{Non-Markovian modeling of Fermi/Boson systems coupled to one or several Fermi/Boson thermal baths}

\author{Denis Lacroix } \email{denis.lacroix@ijclab.in2p3.fr}
\affiliation{Universit\'e Paris-Saclay, CNRS/IN2P3, IJCLab, 91405 Orsay, France}
\author{V.V. Sargsyan}
\affiliation{Joint Institute for Nuclear Research, 141980 Dubna, Russia}
\author{G.G. Adamian}
\affiliation{Joint Institute for Nuclear Research, 141980 Dubna, Russia}
\author{N.V. Antonenko}
\affiliation{Joint Institute for Nuclear Research, 141980 Dubna, Russia}
\affiliation{Tomsk Polytechnic University, 634050 Tomsk, Russia}
\author{A.A. Hovhannisyan}
\affiliation{Joint Institute for Nuclear Research, 141980 Dubna, Russia}
\affiliation{Institute of Applied Problems of Physics, Quantum Computing Laboratory,  0014 Yerevan, Armenia}

\date{\today}
\begin{abstract}
A method is proposed to describe Fermi or Bose systems coupled to one or several heat baths
composed of fermions and/or bosons. The method, called Coupled Equations of Motion method,
properly includes non-Markovian effects. The approach is
exact in the Full-Coupling approximation when only bosonic particles
are present in the system and baths. The approach provides an approximate treatment when fermions are present either in the system and/or
in one or several environments. The new approach has the advantage to properly
respect the Pauli exclusion principle for fermions during the evolution. We illustrate the approach for the single Fermi or Bose two-level system coupled to one
or two heat-baths assuming different types of quantum statistics (Fermion or Bosons) for them.
The cases
of Fermi system coupled to fermion or boson heat baths or a mixture of both are analyzed in details. With the future goal to treat Fermi systems formed of
increasing number of two-level systems (Qubits), we discuss possible simplifications that could be made in the equations of motion and their limits of validity in terms of the system--baths coupling or of the initial heat baths temperatures.
\end{abstract}



\maketitle

\section{Introduction}
The description of dissipation and decoherence is an important subject of investigation,
especially in view of the current boom in quantum technologies \cite{Li20a,Li20b} (see also discussion in Ref. \cite{Mor19}).
In this field, the system of interest is a number of Qubits that are inherently coupled to one or several
environments \cite{Fey82,Bar11,Geo14}. The effect of this coupling is rather dramatic since it induces a non-unitary
evolution of the system, and ultimately tends to destroy the interesting quantum aspects, driving the system to
classical physics.  Controlling the transition from quantum to classical physics is becoming crucial
in this context. By changing the properties of the baths, one might first study the transition
from the Markovian to the non-Markovian regime. In particular, it has been recently underlined that
non-Markovian dynamics and their impact should be explored more systematically \cite{Li20a,Li20b}.

Non-Markovian effects and their description is an active
field of research in the theory of open quantum systems \cite{Riv14,Bre16,Veg17,Li18}, especially with the aim of
developing accurate and versatile approaches.  With the progress of manipulating atoms and molecules, one might
engineer systems and/or environments that can be formed of fermionic or bosonic particles.  Another example is the atomic nuclei where nucleons (fermions) act as a reservoir for the collective excitations treated as bosons \cite{Rin80}.
Treating fermions is finally of special interest in the context of quantum computing due to the one-to-one mapping between spin systems and fermions on lattices \cite{Jor28}.

Our primary goal here is to develop a unified
approach able to describe fermions or bosons coupled to a set of baths that can also be a mixture of fermions and bosons.  System coupled to several reservoirs are of special interest
in different fields of physics. To quote few of them, we mention the example of cavity
quantum electrodynamics  \cite{Maj12}, Jaynes-Cummings lattices \cite{Nun11},
photon-ion interfaces \cite{Lam11}, ion chain systems \cite{Lin11}, or
phonon-induced spin squeezing \cite{Ben13} (see also examples in Refs.~\cite{Che13,Cha14,Hov18,Mwa19}).

In the following, we first recall some recent progress we have made in the description of systems coupled
to several heat baths  \cite{Sar14,Lac15,Sar17,Hov18,Sar18,Hov19,Hov20}. We then discuss in details the extra subtleties
that appear for Fermi systems or environments compared to the Bose case. In the case of bosonic system and environments,
an exact treatment is a priori possible including fully the non-Markovian effects. When fermions are present
in the system or in the surrounding baths, some approximations are required. We describe in section \ref{sec:extfb} a specific
methodology to treat Fermi system coupled to Fermi, Bose or Fermi--Bose mixtures of baths. In this new approach, where the non-Markovian
effects are included, a special attention is paid to respect the
Pauli exclusion principle for Fermi degrees of freedom.
The approach is then illustrated for the system coupled to one or several heat baths
in section \ref{sec:result}.

\section{Method}

We follow here our previous work \cite{Sar14,Lac15,Sar17,Hov18,Sar18,Hov19,Hov20} and consider a Fermi or Bose  system coupled to one or several baths. Some of the baths could be composed of fermions and some other of bosons. In our previous studies we gradually considered  problems of increasing complexity, i.e. changing the quantum statistics of the system or bath, considering more general coupling and/or increasing the number of baths.

A single two-level system coupled to an environment is considered. The system+environment Hamiltonian is taken
as
\begin{eqnarray}
H=H_{\rm S}+H_{\rm E}+H_{\rm SE}. \label{eq:HH}
\label{ham}
\end{eqnarray}
The system and environment Hamiltonians, denoted respectively by $H_S$ and $H_E$, are given by
\begin{eqnarray}
\displaystyle H_{\rm S}=\hbar \omega_1 a^\dagger_1 a_1, ~~~\displaystyle  H_{\rm  E}=\sum_{\nu } \hbar \omega _{\nu }a_{\nu }^\dagger a_{\nu } .
\label{eq:se}
\end{eqnarray}
For the moment, we  do not specify if there is one or several baths and simply assume that the quantum nature (fermionic or bosonic)  of  each pair of creation/annihilation operators $(a^\dagger_\nu, a_\nu)$ is specified through the relation
\begin{eqnarray}
a_\nu a^\dagger_\nu = 1+ \varepsilon_\nu a^\dagger_\nu a_\nu ,
\end{eqnarray}
where $\varepsilon_\nu = +1$ ($-1$) for bosons (fermions). We also use the convention $\nu=1$ ($\nu > 1$) for the system
(for the environment).
Note that, in the present model, fermionic heat-bath is described by an infinite set of two-level systems initially at thermal equilibrium. Besides the commutation/anti-commutation rules of creation/annihilation operators, the difference between a bosonic and fermionic bath stems from the initial occupation probability
denoted by $n(\omega_\nu)$
that corresponds either to Bose-Einstein or Fermi-Dirac occupation probability. A pictorial view of the heat-bath was given in Fig. 1 of Ref. \cite{Sar14}. 

In the following, we will consider
the Full--Coupling case (FC), in which the coupling between the system and environment is as follows
\begin{eqnarray}
H_{\rm SE} &=& H_{\rm FC} = \sum _{\nu>1} g_{\nu } (a^\dagger_1 +a_1 )\left(a_{\nu }^\dagger +a_{\nu }\right). \label{eq:fc}
\end{eqnarray}
The present formulation is rather flexible and includes the possibility that some environmental particles obey fermion statistics
while other obey boson statistics. It also includes the possibility that the environment can be decomposed into several baths with eventually different quantum natures and different initial temperatures.  We note that the Hamiltonian (\ref{eq:HH}) can be used
to describe a single Qubit surrounded by one or several heat-baths. In our previous
work, we used the Heisenberg representation to obtain a solution to the system+environment problem taking into account possible non--Markovian effects \cite{Sar14,Lac15,Sar17,Hov18,Sar18,Hov19,Hov20}.  The solution we provided, once the frequency is properly renormalized (see discussion below), is exact for the 
FC coupling when only bosonic particles are considered for both the system and the heat-baths. The situation is more delicate in the FC case when fermions (either in the bath and/or one of the environments) are involved. In this case, the problem could not be solved exactly and a specific prescription should be made to obtain a closed form of
the equations of motion to be solved. A first solution to this problem was given in Ref. \cite{Sar17,Sar18}.  Such a solution
resulted in quite reasonable description of the Fermi systems coupled to one or several baths. Numerical applications have
recently shown however that occupation numbers of the systems with the fermionic heat bath(s) can sometimes slightly exceed $1$. This points out that some modifications of the method might be needed.
Another interesting result was the absence
of asymptotic stationary solution when the system is coupled to a mixture of fermionic and bosonic heat-baths \cite{Hov19}.
In the following, we propose an alternative solution that avoids the occurrence of non-physical occupation numbers during the evolution.

\subsection{Summary and illustration of our previous work}

We summarize here the strategy we employed previously starting from the Heisenberg equations of motion for the system and heat-bath creation operators:
\begin{eqnarray}
\displaystyle \frac{d}{dt}a^{\dag}_1 &=&i\omega_1 a^{\dag}_1 +i(1-[1-\varepsilon_1]a^{\dag}_1a_1)   \sum_{\nu} g_{\nu} \left[ a_{\nu}^{\dag}+ a_{\nu} \right],   \label{eq:eom1.1} \\
\displaystyle  \frac{d}{dt}a^{\dag}_{\alpha}&=&i\omega_{\alpha} a^{\dag}_{\alpha} +i g_{\alpha} (1-[1-\varepsilon_\alpha] a^{\dag}_{\alpha} a_{\alpha})[  a^{\dag}_1 +  a_1]. \label{eq:eom1.2} 
\end{eqnarray}
When all particles are bosons, we have $[1-\varepsilon_1]=[1-\varepsilon_\alpha]=0$ for all $\alpha>1$. The equations of motion become a linear set of equations between the creation/annihilation operators. In this case, the problem is solved exactly  using the Laplace transform technique (see for instance \cite{Sar14}).
In Refs.~\cite{Lac15,Sar17}, we have also illustrated that such a problem can be accurately solved by using the discretized environment method (DEM) together with the special Bogolyubov transformation between the creation/annihilation operators (see section IV of Ref. \cite{Sar17}). An alternative solution to the present problem could be to consider directly the coupled equations of motion (CEM) for the normal  and anomalous densities, denoted respectively by $M$ and $K$, associated to the system+environment.
Using the notations \footnote{Note that for the sake of simplicity, we use slightly different convention for the indices ordering compared to the standard definition of the normal and anomalous densities in many-body systems \cite{Rin80}.}:
\begin{eqnarray}
\left\{
\begin{array} {l}
M_{\nu\alpha} =  \langle a^\dagger_\nu a_\mu \rangle , ~~~G_{\nu \alpha}(t)  =   \langle a_\nu a^\dagger_\alpha \rangle, \\
\\
K_{\nu \alpha} = \langle a^\dagger_\nu a^\dagger_\alpha  \rangle , ~~~K^*_{\nu \alpha}  = \langle a_\alpha a_\nu \rangle,
\end{array}
\right.
\label{eq:dk}
\end{eqnarray}
we obtain for the boson system coupled to bosonic environment
the set of coupled equations for the $M$ and $K$ components:
\begin{widetext}
\begin{eqnarray}
\left\{
\begin{array} {ll}
\displaystyle \frac{d M_{11}}{dt} &=  i \sum_\nu g_\nu (M_{\nu 1}  - M_{1\nu}) +  i \sum_\nu g_\nu (K^*_{1\nu} -
K_{1\nu} )   \\
\\
\displaystyle
 \frac{dM_{1\alpha} }{dt} &=      i  (\omega_1 - \omega_\alpha) M_{1\alpha}
+i  \sum_{\nu} g_{\nu} (M_{\nu\alpha} + K^*_{\alpha\nu})
-i g_\alpha (M_{11} + K_{11})  \\
\\
\\
\displaystyle \frac{d M_{\alpha \beta} }{dt} &=   i  (\omega_\alpha - \omega_\beta )M_{\alpha \beta}
+ i g_{\alpha}  (M_{1\beta} + K^*_{\beta 1})
-i g_\beta   (M_{\alpha 1} + K_{\alpha 1}) \\
\\
\displaystyle  \frac{d K_{11} }{dt} &= 2i  \omega_1 K_{11} +i \displaystyle  \sum_{\nu} g_{\nu} ( K_{\nu 1} + M_{1\nu} + G_{\nu 1} + K_{1\nu})  \\
\\
\displaystyle\frac{d K_{1\alpha}}{dt} &= i  (\omega_1 + \omega_\alpha)  K_{1\alpha}
+i    \sum_{\nu} g_{\nu} (G_{\nu \alpha} + K_{\nu\alpha})  +i  g_\alpha   (M_{11}+K_{11}) \\
\\
\displaystyle \frac{d K_{\alpha\beta}  }{dt}&= i  (\omega_\alpha + \omega_\beta)  K_{\alpha\beta}
+ i g_\alpha (G_{1 \beta} + K_{1\beta})
 + i g_\beta (M_{\alpha 1} + K_{\alpha1}).
\end{array}
\right.   \label{eq:sebosons}
\end{eqnarray}
\end{widetext}
Solving these equations
numerically, we obtain the exact solution of the problem when only bosons are presented in both the system and baths. As illustrated below, this method
is strictly equivalent to the one we used in Refs.~\cite{Sar14,Lac15}.

 \begin{figure}[!h]
    \includegraphics[width=1.0 \linewidth]{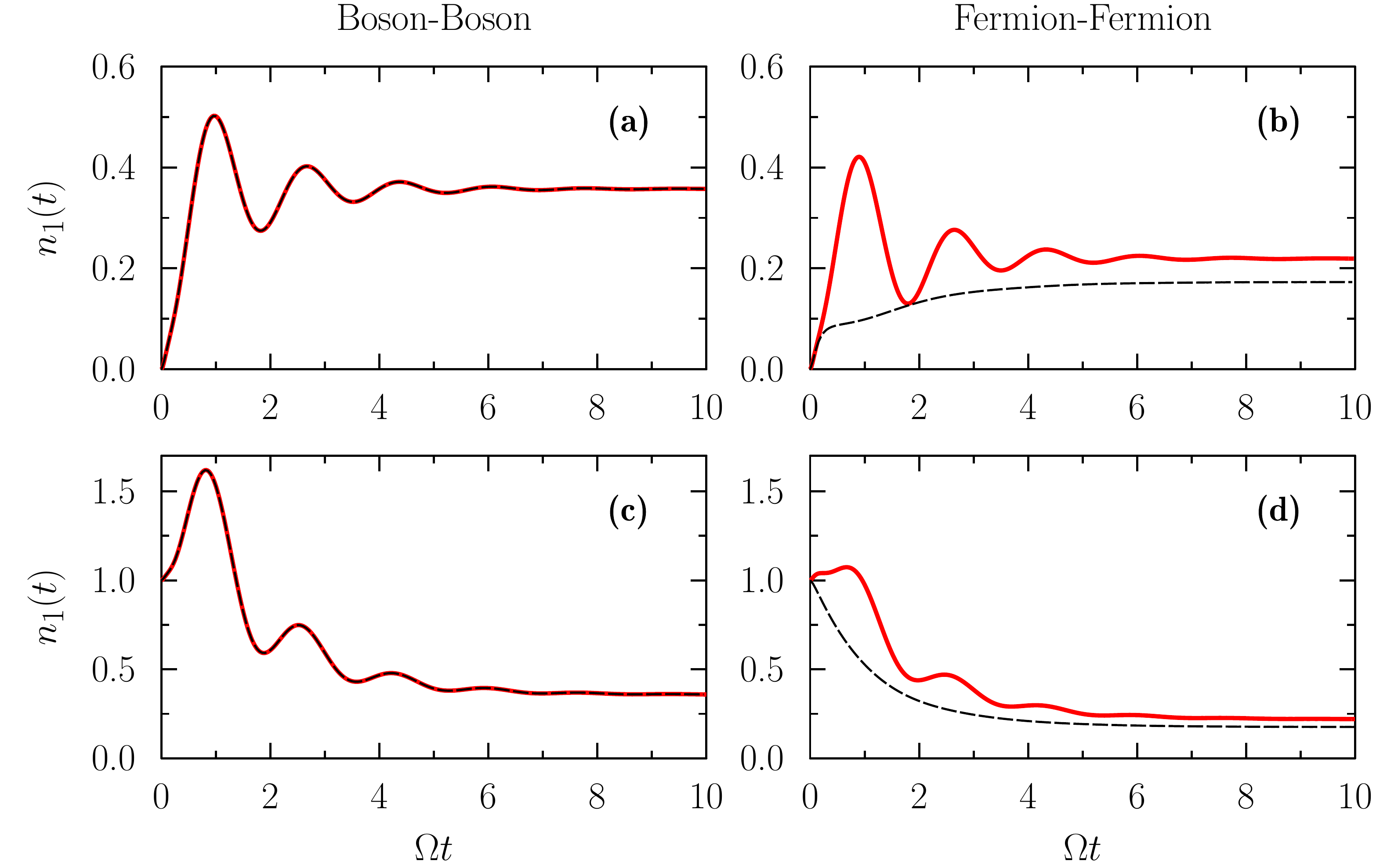}
    \caption{Time evolution of the occupation probability in the system $n_1(t)=M_{11}(t)$ obtained by solving Eqs.~(\ref{eq:sebosons}) (red solid line) or by solving the set of equations  (\ref{eq:fcfulllast}) (black dashed line).
    Panels (a) and (c) correspond to the case of a bosonic system coupled
    to a bosonic bath starting from  $n_1(0) = 0$ and  $n_1(1) = 1$ respectively.
    Panels (b) and (d) correspond to the case of a fermionic system coupled to a fermionic bath again starting from $n_1(0) = 0$ and  $n_1(1) = 1$
    respectively. In all cases, the bath properties are given by $c_1=0.1$,
    $\gamma_1/\Omega=12$ and $T_1/\Omega=1$. Note that, in the B-B case, the two approaches are equivalent and the two curves displayed
    in panels (a) and (c) cannot be distinguished. }
    \label{fig:BBFFold}
\end{figure}


\subsection{Numerical solution of Eq.~(\ref{eq:sebosons}) with discretization of heat bath}
\label{sec:dem}

Here, we consider the system with frequency $\Omega$ (we use the convention $\hbar=1$ thorough the paper) and all energies,
coupling constants are given in units of $\Omega$. Time  is given in $\Omega^{-1}$ units. We assume that the system is coupled to one or several baths. Each bath, labelled by $i$ has a Lorentz-Drude spectral function given by
\begin{eqnarray}
J_i(\omega) &=&  \frac{c_i}{\pi}  \omega \frac{\gamma^2_i}{\gamma^2_i + \omega^2}. \label{eq:lorentz}
\end{eqnarray}
The two parameters $c_i$ and $\gamma_i$ determine the coupling strength with the system and the memory effect respectively.
In order to solve the CEM,
a finite discrete number of levels $\nu$ is used for each environment.  In practice, we follow the procedure proposed in Ref.  \cite{Lac15}. For a given environment, a finite set of frequencies $\omega_\nu$ is used according to:
\begin{eqnarray}
\omega_\nu & = & \Delta \omega (n + 1/2), ~~n=0,\cdots,N_{\rm max}. \label{eq:dw0}
\end{eqnarray}
Then, the coupling $g_\nu$ entering in the Hamiltonian is given by $g_\nu =   \sqrt { \Delta \omega J_i (\omega_\nu) }$. Further details and discussions can be found in Ref. \cite{Lac15}. Note that the discretization of $\omega$ can be taken non-uniform. In practice, small $\Delta \omega$ are required only in the vicinity of $\Omega$ to get a good numerical accuracy and larger $\Delta \omega$ can be used away from the system frequency. We use this property to reduce the total number of states
necessary for each bath. In the following applications,
we use $N_{\rm max}=400$ states with $\Delta \omega$ that depends on $n$ such that
$\Delta \omega_{n+1} = \lambda \Delta \omega_{n}$ with $ \Delta \omega_{0}/\Omega = 0.01$  and $\lambda=1.015$.
The choice of the values for three parameters $\Delta \omega_{0}$, $\lambda$ and $N_{\rm max}$ is critical to properly achieve good numerical accuracy. In particular, these parameters should be chosen in such a way that $\Delta \omega_{n} \ll \Omega$ in the vicinity of $\Omega$ and $T \ll \omega_{N_{ max}}$ while keeping $N_{\rm max}$ not too high in order to obtain the result in a reasonable numerical time. The choice of the parameters for the discretization has been validated for the boson system coupled to bosonic bath by comparing with the results obtained using the Laplace transform technique \cite{Sar14}.


To avoid the unphysical shift of the system frequency induced by the coupling with the baths, this frequency is renormalized
prior to the calculation as it is always done \cite{Bre02}. For
the FC case, the frequency $\omega_1$
used in Eqs.~(\ref{eq:sebosons}) is as follows
\begin{eqnarray}
\omega_1 = \Omega +4  \sum_\nu \frac{g_\nu^2}{\omega_\nu}.
\end{eqnarray}
Finally, each heat-bath is characterized by its initial temperature $T_i$ such that the initial occupation $n_\nu (0)$ of the state in the bath $i$ is given by (using the convention $k_B=1$ for the Bolztmann constant):
\begin{eqnarray}
n_\nu(0) & = & M_{\nu \nu} (0) =  \frac{1}{\exp\left(\omega_\nu / T_i  \right) - \varepsilon_\nu}.
\end{eqnarray}

The results are shown in Fig.~\ref{fig:BBFFold} for the Bose system coupled to single bosonic bath.
The time evolution of occupation probability is obtained by solving  the set of equations (\ref{eq:sebosons}) for the FC coupling using the discretization of the environment. The results perfectly match those obtained in Ref.~\cite{Sar14} using the Laplace transform technique. When we write down the creation/annihilation operators $a_1^+$/$a_1$
($a^{\dagger}_{\nu}/a_{\nu}$), we mean the creation/annihilation
operators of transition with the corresponding energy $\hbar \omega_1$
($\hbar \omega_{\nu}$). So, each $a_1$ ($a^{\dagger}_{\nu}$) and $a_1$
($a_{\nu}$) is the product of creation and annihilation operators of
particle. In our formalism, there is a conversion of excitation quanta
from the fermionic or bosonic system to the bosonic or fermionic environment
or vice versa. Since only the level associated to $a^\dagger_1$ is populated or depopulated  by the coupling, the number of particles in the system as a function of time directly identifies with the quantity $n_1(t)$. 

\subsection{Direct extension of Eqs.~(\ref{eq:sebosons}) when fermions are involved}

In the following, we will systematically use the short-hand notation B-B, F-F, B-F, and F-B where the first letter refers to the system statistics (B=Bosons and F=Fermions) while the second letter refers to the bath statistics.  When more than one bath is considered, we use the convention
System-Bath$_1$-Bath$_2$-...

Equations ~(\ref{eq:sebosons}) provide an exact treatment of the B-B case.  Based on simple arguments, we previously proposed to treat the F-F case by neglecting the terms $(1-\varepsilon_1) a^\dagger_1 a_1 $  and $(1-\varepsilon_\alpha) a^\dagger_\alpha a_\alpha$  in Eqs.~(\ref{eq:eom1.1}) and (\ref{eq:eom1.2}) when the system and baths are both composed of fermions. This direct mapping from bosons to fermions has the great advantage to give linear Heisenberg equations also for the F-F case while the bath properly imposes asymptotically the Fermi statistics to the system in the weak coupling--high temperature limit.
The results of this approximate treatment are also shown in Fig.  \ref{fig:BBFFold} for the F-F case. We clearly see in this figure (panel (d)) that, while the asymptotic behavior is expected
to be properly treated, the price to pay with the simplified treatment is the occurrence of unphysical behavior
at initial time-scale with occupation numbers larger than $1$. This stems from the fact that the Pauli exclusion
principle might be broken during the time evolution. Indeed, in Eqs.~(\ref{eq:sebosons}), nothing
prevents from having $K_{11}(t) = \langle (a^\dagger_1)^2 \rangle$ non-zero during the evolution even if the system is
fermionic.

\begin{figure}[!h]
    \includegraphics[width=1.0 \linewidth]{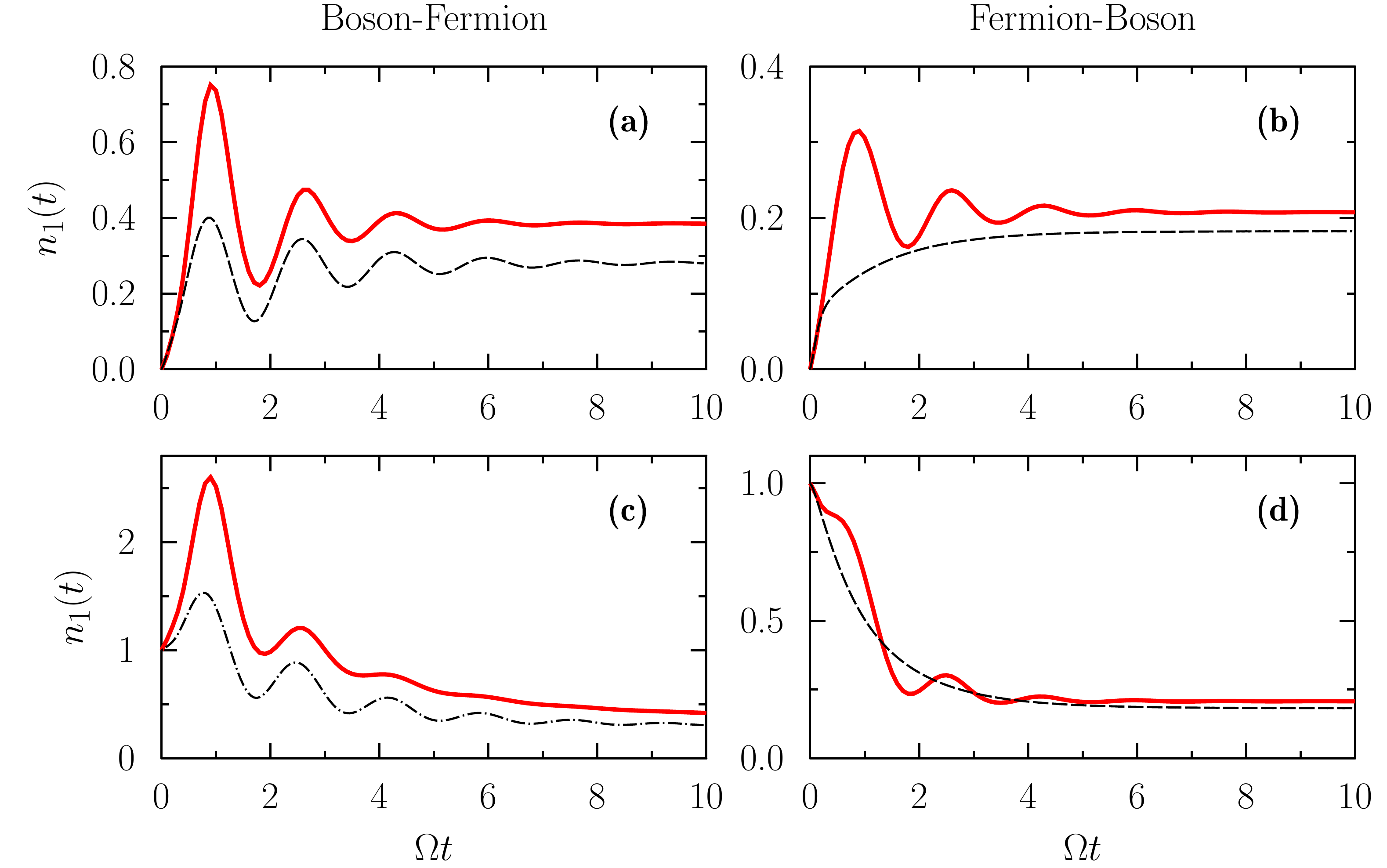}
    \caption{The same as in Fig.~\ref{fig:BBFFold}, but for the mixed quantum statistics case of the Boson system coupled to Fermion bath (B-F) [(a) and (c)] and for the Fermion system coupled to Boson bath (F-B) [(b) and (d)].
 The bath properties are given by $c_1=0.1$, $\gamma_1/\Omega=12$ and $T_1/\Omega=1$. In all cases, the red solid line corresponds to the result obtained with the Laplace transform approach Ref.~\cite{Sar18}. The black dashed line corresponds to the result obtained with Eqs.~(\ref{eq:fcfulllast}).
    }
    \label{fig:BFFBold}
\end{figure}

Nevertheless, the simplified treatment has additional interesting properties that have been used to overcome this difficulty.
One of them is the possibility to map exactly  Eqs.~(9) into the simple time-local diffusion equation for the occupation probability:
\begin{eqnarray}
\frac{d n_1(t)}{dt} &=& -2 \lambda_1 (t) n_1(t) + 2D_1(t) , \label{eq:diffusion}
\end{eqnarray}
including fully non-Markovian effects. Using this equation and some symmetry properties of the master
equation obtained for $n_1(t)$ with the non-linear termed, it was shown in Ref.~\cite{Sar18}   that
the F-B and B-F case can be accurately described using the diffusion equation (\ref{eq:diffusion}) with modified
transport coefficients. This approach leads to the proper asymptotic limit even though the Fermi nature of the system
or bath might be slightly broken. The results obtained with this method for the F-B and B-F case are illustrated in Fig.~\ref{fig:BFFBold}.

\section{Treatment of F-F, B-F, and F-B cases enforcing the Pauli exclusion principle}
\label{sec:extfb}

In the present paper we propose a treatment that respects the Fermi nature of the particle all along the non-equilibrium evolution. This implies to explicitly account for the non-linear term in
the Heisenberg equation of motion.  The situation is similar
to the many-body problem of interacting particles where the one-body density matrix evolution
depends on the two-body density, whose evolution is itself coupled to the three-body density and so on and so forth, leading
to the so-called Bogolyubov- Born-Green-Kirkwood-Yvon hierarchy \cite{Bog46, Bor46, Kir46,Cas90,Gon90,Sch90,Bon16}.

The general form of the equations of motion, which is valid regardless of the quantum natures of the system or bath, is given as
\begin{widetext}
\begin{eqnarray}
\left\{
\begin{array}{ll}
\displaystyle
\frac{d \langle a^\dagger_1 a_1 \rangle }{dt} &=  i \sum_\nu g_\nu (\langle a^\dagger_\nu a_1 \rangle  - \langle a^\dagger_1 a_\nu \rangle) +  i \sum_\nu g_\nu ( \langle a_\nu a_1 \rangle - \langle a^\dagger_1 a^\dagger_\nu \rangle )  \\
\\
\displaystyle \frac{d \langle a^\dagger_1 a_\alpha \rangle }{dt} &=    i  (\omega_1 - \omega_\alpha) \langle a^\dagger_1 a_\alpha \rangle
+i   \sum_{\nu} g_{\nu} \langle (1-[1-\varepsilon_1]a^{\dag}_1a_1)  \left[ a_{\nu}^{\dag} a_\alpha + a_{\nu} a_\alpha \right] \rangle
-i g_{\alpha} \langle (1-[1-\varepsilon_\alpha] a^{\dag}_{\alpha} a_{\alpha}) [a^{\dag}_1 a_1 + a^{\dag}_1 a^\dagger_1]\rangle \\
\\
\displaystyle \frac{d \langle a^\dagger_\alpha a_\beta \rangle }{dt} &=   i  (\omega_\alpha - \omega_\beta )\langle a^\dagger_\alpha a_\beta \rangle
+ i g_{\alpha}\langle  (1-[1-\varepsilon_\alpha] a^{\dag}_{\alpha} a_{\alpha})[  a^{\dag}_1 a_\beta +  a_1 a_\beta ] \rangle
-i g_\beta\langle [a^\dagger_\alpha a^\dagger_1 + a^\dagger_\alpha a_1 ] (1-[1-\varepsilon_\beta] a^{\dag}_{\beta} a_{\beta}) \rangle \\
\\\displaystyle
\frac{d \langle a^\dagger_1 a^\dagger_1 \rangle}{dt} &= 2i  \omega_1  \langle a^\dagger_1 a^\dagger_1 \rangle
+i\displaystyle \frac{1+\varepsilon_1}{2}   \sum_{\nu} g_{\nu}\left( \langle  (1-[1-\varepsilon_1]a^{\dag}_1a_1) [ a_{\nu}^{\dag}a^\dagger_1 + a_{\nu} a^\dagger_1 ]\rangle
+ \langle  [ a^\dagger_1 a_{\nu}^{\dag} + a^\dagger_1a_{\nu} ] (1-[1-\varepsilon_1]a^{\dag}_1a_1) \rangle \right) \\
\\
\displaystyle
\frac{d \langle a^\dagger_1 a^\dagger_\alpha \rangle}{dt} &= i  (\omega_1 + \omega_\alpha) \langle a^\dagger_1 a^\dagger_\alpha \rangle
+i   \sum_{\nu} g_{\nu}\langle  (1-[1-\varepsilon_1]a^{\dag}_1a_1) \left[ a_{\nu}^{\dag}a^\dagger_\alpha + a_{\nu} a^\dagger_\alpha \right]\rangle
+ i g_{\alpha} \langle [a^\dagger_1  a_1 + a^\dagger_1a^\dagger_1] (1-[1-\varepsilon_\alpha] a^{\dag}_{\alpha} a_{\alpha})\rangle   \\
\\
\displaystyle
\frac{d \langle a^\dagger_\alpha a^\dagger_\alpha \rangle}{dt} &= 2i  \omega_\alpha  \langle a^\dagger_\alpha a^\dagger_\alpha \rangle
+i \displaystyle\frac{1+\varepsilon_\alpha}{2}   g_{\alpha}\left( \langle  a^{\dag}_\alpha a_1 \rangle
+ \langle  a^\dagger_\alpha a_{1}^{\dag}\rangle + \langle a_1 a^\dagger_\alpha \rangle + \langle a^{\dag}_1a^{\dag}_\alpha \rangle \right) \\
\\
\displaystyle
\frac{d \langle a^\dagger_\alpha a^\dagger_\beta  \rangle}{dt} &= i  (\omega_\alpha + \omega_\alpha)
 \langle a^\dagger_\alpha a^\dagger_\beta  \rangle + i g_{\alpha} \langle (1-[1-\varepsilon_\alpha] a^{\dag}_{\alpha} a_{\alpha})[  a^{\dag}_1 a^\dagger_\beta  +  a_1 a^\dagger_\beta] \rangle + i g_{\beta} \langle [  a^\dagger_\alpha a^{\dag}_1  +  a^\dagger_\alpha a_1]  (1-[1-\varepsilon_\beta] a^{\dag}_{\beta} a_{\beta})\rangle .
 \end{array}
\right.   \label{eq:fullexact}
\end{eqnarray}
\end{widetext}
When the system and environment contain only bosons, we have $[1-\varepsilon_1] = [1-\varepsilon_\alpha] = 0$ for all $\alpha$ and we
recover the set of equations (\ref{eq:sebosons}). The evolution of the occupation number $M_{11}$  depends on the off-diagonal elements
$M_{1\nu}$, $M_{\nu1}$, $K_{1\nu}$, and $K^{*}_{1\nu}$
of the normal and anomalous densities whose evolutions
depend on $\langle a^{\dag}_1a_1 a_{\nu}^{\dag}a_\alpha \rangle    $, $\langle a^{\dag}_1a_1 a_{\nu}a_\alpha \rangle $,
$\langle a^{\dag}_{\alpha} a_{\alpha} a^\dagger_1 a^{\dag}_1 \rangle$, $\langle a^{\dag}_{\alpha} a_{\alpha} a^\dagger_1 a_1 \rangle $, ...
These degrees of freedom are themselves coupled to higher-order moments related to higher-order quantum fluctuations.
The full problem cannot be solved exactly, due to the number of degrees of freedom
that should be followed in time when fermions are considered. 

If the system is driven by the fermionic and/or bosonic harmonic potentials that destroy high-order quantum fluctuations,
then these fluctuations are presented as a product of two diagonal elements or a product of diagonal
($M_{11}$, $M_{\alpha\alpha}$, $K_{11}$, $K_{\alpha\alpha}$)
and off-diagonal ($M_{\nu \alpha}$, $K^*_{\alpha \nu}$, $K_{\nu\alpha}$, where $\alpha\ne\nu$) elements of the normal and anomalous densities,
  \begin{eqnarray}
  \left\{
  \begin{array} {l}
\displaystyle \langle a^{\dag}_1a_1 a_{\nu}^{\dag}a_\alpha \rangle
\simeq \langle a^{\dag}_1a_1  \rangle \langle a_{\nu}^{\dag}a_\alpha \rangle                = M_{11} M_{\nu \alpha}   \\
\\
\langle a^{\dag}_1a_1 a_{\nu}a_\alpha \rangle
\simeq \langle a^{\dag}_1a_1  \rangle \langle a_{\nu}a_\alpha \rangle                       = M_{11} K^*_{\alpha \nu} \\
\\
\langle a^{\dag}_{\alpha} a_{\alpha} a^\dagger_1 a^{\dag}_1 \rangle
\simeq \langle a^{\dag}_{\alpha} a_{\alpha} \rangle \langle  a^\dagger_1 a^{\dag}_1 \rangle = M_{\alpha\alpha} K_{11} \\
\\
\langle a^{\dag}_{\alpha} a_{\alpha} a^\dagger_1 a_1 \rangle
\simeq \langle a^{\dag}_{\alpha} a_{\alpha} \rangle \langle  a^\dagger_1 a_1 \rangle        = M_{\alpha\alpha} M_{11}\\
\\
\cdots
\end{array}
\right. \label{eq:moments}
\end{eqnarray}
This truncation procedure is equivalent to the linearization of the equations of motion with respect to the off-diagonal elements
$M_{\alpha\beta}$, $K_{\alpha\beta}$,  $K^{*}_{\alpha\beta}$
($\alpha\ne\beta$, including the cases of $\alpha=1$, $\beta=1$).
The right hand sides of these equations contain the terms with the off-diagonal elements
only in the first order, i.e., the
terms containing product  of two off-diagonal elements are neglected,
since   they are very small with respect
to the corresponding terms  that are proportional to a product of two diagonal elements or a product of diagonal and off-diagonal  elements.


Employing the mean-field type approximation (\ref{eq:moments}), we obtain from the
exact Eqs. (\ref{eq:fullexact}) the closed set of equations of motion:
\begin{widetext}
\begin{eqnarray}
\left\{
\begin{array} {ll}
\displaystyle \frac{d M_{11}}{dt} &=  i \sum_\nu g_\nu (M_{\nu 1}  - M_{1\nu}) +  i \sum_\nu g_\nu (K^*_{1\nu} -
K_{1\nu} )   \\
\\
\displaystyle
 \frac{dM_{1\alpha} }{dt} &=      i  (\omega_1 - \omega_\alpha) M_{1\alpha}
+i \xi_1 (t) \sum_{\nu} g_{\nu} (M_{\nu\alpha} + K^*_{\alpha\nu})
-i g_\alpha \xi_\alpha(t)  (M_{11} + K_{11})  \\
\\
\displaystyle  \frac{d M_{\alpha\alpha}}{dt} &=  i g_\alpha (M_{1\alpha}  - M_{\alpha 1} + K^*_{\alpha 1} - K_{\alpha 1}) \\
\\
\displaystyle \frac{d M_{\alpha \beta} }{dt} &=   i  (\omega_\alpha - \omega_\beta )M_{\alpha \beta}
+ i g_{\alpha} \xi_\alpha(t) (M_{1\beta} + K^*_{\beta 1})
-i g_\beta \xi_\beta(t)  (M_{\alpha 1} + K_{\alpha 1}) \\
\\
\displaystyle  \frac{d K_{11} }{dt} &= 2i  \omega_1 K_{11} +i \displaystyle \frac{(1+ \varepsilon_1)}{2} \sum_{\nu} g_{\nu} ( K_{\nu 1} + M_{1\nu} + G_{\nu 1} + K_{1\nu})  \\
\\
\displaystyle\frac{d K_{1\alpha}}{dt} &= i  (\omega_1 + \omega_\alpha)  K_{1\alpha}
+i  \xi_1(t)   \sum_{\nu} g_{\nu} (G_{\nu \alpha} + K_{\nu\alpha})  +i  g_\alpha   \xi_{\alpha} (t) (M_{11}+K_{11}) \\
\\
\displaystyle \frac{d K_{\alpha\alpha}  }{dt}&= 2 i  \omega_\alpha K_{\alpha\alpha} +\displaystyle  i \frac{(1+\varepsilon_\alpha)}{2} g_\alpha \left( M_{\alpha 1} + K_{\alpha 1} + G_{1 \alpha} + K_{\alpha 1}\right) \\
\\
\displaystyle \frac{d K_{\alpha\beta}  }{dt}&= i  (\omega_\alpha + \omega_\beta)  K_{\alpha\beta}
+ i g_\alpha \xi_\alpha(t) (G_{1 \beta} + K_{1\beta})
 + i g_\beta \xi_\beta(t) (M_{\alpha 1} + K_{\alpha 1}),
\end{array}
\right.   \label{eq:fcfulllast}
\end{eqnarray}
\end{widetext}
where $\xi_\alpha(t)  =  \langle [a_\alpha, a^\dagger_\alpha] \rangle$ (including the case of $\alpha$=1).
This set of equations is the main result of the present work.
It can be applied regardless of the quantum natures of the system or baths (fermionic or bosonic)
as well as to system coupled to several baths.  In the following, the method will be called Coupled Equations of Motion (CEM) method.

There are a number of properties of $M$, $G$, and $K$ components
which help us to solve these equations.  Assuming that each bath is composed of a set of independent two-level
systems, we have for all $\alpha$ and $\beta$:
\begin{eqnarray}
K_{\alpha \beta} &=& K_{\beta\alpha}, \nonumber \\
G_{\alpha \beta} &=& M_{\beta\alpha}, ~~~(\alpha \neq \beta)   \nonumber \\
G_{\alpha\alpha} &=& 1 + \varepsilon_\alpha M_{\alpha \alpha}.  \nonumber
\end{eqnarray}
The last relation implies that we also have $\xi_\alpha(t)  =  1 + [\varepsilon_\alpha -1] M_{\alpha\alpha}$.

When the system and baths are all composed of bosons, we have   $\xi_1(t)= 1$ and  $\xi_\alpha(t) = 1$ for all $\alpha$
and it could be easily shown that Eqs.~(\ref{eq:fcfulllast}) are reduced to Eqs.~(\ref{eq:sebosons}). Therefore we also
obtain an exact solution of the problem in the B-B case.
Another important property visible in  (\ref{eq:fcfulllast}) is that, if we assume $\alpha$ to be fermionic, we have simply $\dot K_{\alpha\alpha} = 2i\omega_\alpha K_{\alpha\alpha}$.  Since at initial time $K_{\alpha\alpha}(0)=0$ for fermions, this property is also respected at all time as it should be.

The truncation procedure leading to the set of coupled equations (\ref{eq:fcfulllast}) was tested for the F-F, F-B, and B-F cases. We
found that, for these systems, the occupation numbers $M_{11}$ calculated with (\ref{eq:fcfulllast}) and within the  Langevin
approach of Refs. \cite{Sar14,Lac15,Sar17,Hov18,Sar18,Hov19,Hov20}, taking into account the Pauli principle,  have almost the same time dependencies and asymptotic values.
This indirectly justifies our truncation procedure.

\section{Applications}
\label{sec:result}

\subsection{System coupled to single heat bath}

In the following, the results obtained for the FC Hamiltonian
using the CEM approach for a system coupled to one bath
with various quantum statistics are compared with our previous calculations in Fig. \ref{fig:BBFFold}
for the B-B and F-F and in Fig.~\ref{fig:BFFBold} for the B-F and F-B cases.

There are a number of remarks that can be
made from the comparison. Since Eqs.~(\ref{eq:fcfulllast}) are identical to Eqs.~(\ref{eq:sebosons}) in the B-B case,
we obviously observe in Fig. \ref{fig:BBFFold}  (panels (a) and (c)) that results of the two sets  of equations coincide.
In all cases, we see in Figs. \ref{fig:BBFFold}  and  \ref{fig:BFFBold}
that the time-scale to reach the asymptotic equilibrium is compatible with our previous estimates. However,
when fermions are present in the system and/or bath, the amplitudes of oscillation during the descent to equilibrium are reduced in the new
approach. This is particularly visible when the system is fermionic where the oscillations completely disappear. The asymptotic limit
is more modified  in the F-B or B-F case (Fig.  \ref{fig:BFFBold}) and to a lesser extent in the F-F case (Fig. \ref{fig:BBFFold}).
When the system is fermionic, we see by comparing  Figs.  \ref{fig:BBFFold}
and  \ref{fig:BFFBold} that the quantum nature (fermionic or bosonic) of the bath affects much
less the evolution compared to the case of a bosonic system. This will be systematically observed in all illustrations given below.
Another generic feature is the absence of unphysical values for the occupation probabilities of fermionic system when Eqs. (\ref{eq:fcfulllast})
are solved. This gives indirect indication that the fermionic nature of the system is properly accounted for.

\subsection{Toward simplified  treatments of fermionic or bosonic system coupled to single heat bath}
\label{sec:appfermi}

We propose here a method to treat the Fermi (or Bose) systems coupled to one or several heat-baths
by solving discretized versions of different heat-baths together with numerical integration of a closed set of equations
between the normal and anomalous densities of the system+environment. The discretization technique can be rather costly numerically.
With the aim to treat more elaborated systems (with many Qubits) coupled to many heat-baths, we further explore  the possibility to obtain a simpler framework compared
to Eqs.~(\ref{eq:fcfulllast}).

\begin{figure}[!h]
\includegraphics[width= 1.0\linewidth]{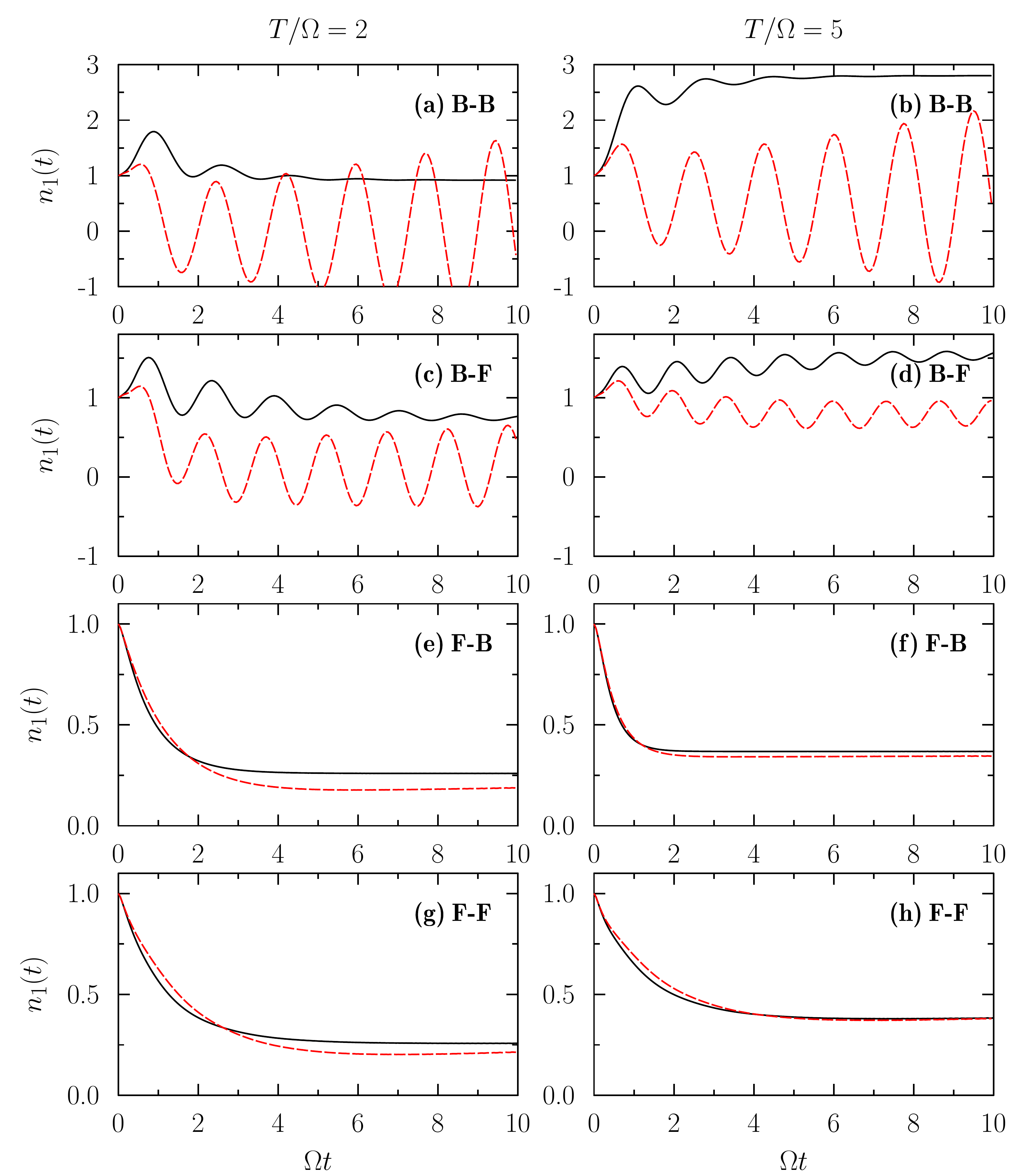}
\caption{Comparison of the evolution of  $n_1(t)$ in the FC coupling for a system coupled to a single bath
obtained  within the CEM approach (black solid lines) at $T_1/\Omega= 2$ (left panels) and $T_1/\Omega= 5$ (right panels). Results are
systematically compared to the CEM approach  assuming $K_{\alpha\beta} = 0$  and $M_{\alpha \beta}=0$ for $\alpha \neq \beta$  during the evolution (red dashed line).
Here, $\alpha$ and $\beta$ belong to the bath.
 The combinations of the system and heat-bath are indicated in each panel.
 All calculations are performed at $\gamma_1/\Omega=12$ and $c_1=0.1$. }
    \label{fig:alldiag}
\end{figure}

One possibility is to 
assume that the off-diagonal matrix elements $M_{\alpha\beta}$ and $K_{\alpha\beta}$ in Eqs.~(\ref{eq:fcfulllast}) are zero when both $\alpha$ and $\beta$
belong to the bath.  This approximation, that is expected to be accurate in the weak-coupling regime, is discussed in more details in Appendix \ref{sec:approx}. In
particular, a connection with the diffusion equation (\ref{eq:diffusion}) is made.
We compare in Fig.   \ref{fig:alldiag} the full and approximate treatments for the F-F, F-B, B-F, and B-B cases. We clearly
see from this figure that the approximation leads to unphysical occupation numbers for  bosonic system.
Surprisingly enough, for Fermi systems, although not perfect, the approximation turns out to reproduce quite well the full evolution
whatever is the nature of bath. We finally also checked numerically for the F-F case that the approximation is better when the temperature increases but
degrades when the coupling strength increases.  The same conclusion can be drawn for a Fermi system coupled to several baths
independently of the quantum nature of the baths.

\begin{figure}[!h]
\includegraphics[width= 1.0\linewidth]{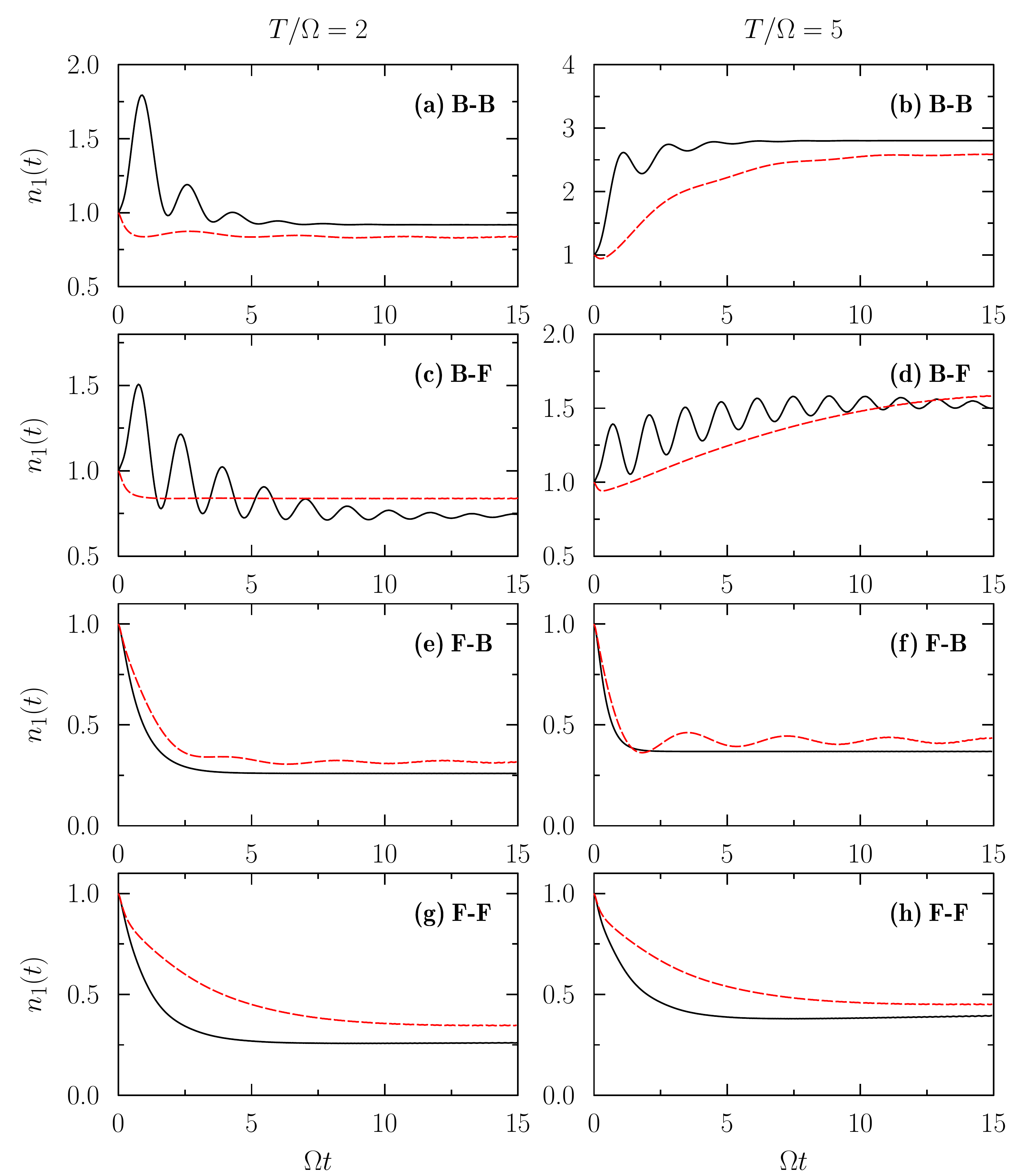}
\caption{Comparison of the evolution of  $n_1(t)$ in the FC coupling for the system coupled to single bath
obtained  within the CEM approach (black solid lines) at $T_1/\Omega= 2$ (left panels) and $T_1/\Omega= 5$ (right panels). Results are
systematically compared to the CEM approach  assuming that the anomalous density is zero ($K = 0$) together with
$M_{\alpha \beta}=0$ for $\alpha \neq \beta$  during the evolution (red dashed line). Here, $\alpha$ and $\beta$ belong to the bath.
 The combinations of the system and heat-bath are indicated in each panel.
 All calculations are performed at $\gamma_1/\Omega=12$ and $c_1=0.1$. }
    \label{fig:alldiagBBBF}
\end{figure}

We explored the possibility to obtain the alternative  simplified description of bosonic system coupled to a bosonic or fermionic
systems. For the bosonic case, one cannot set only part of the anomalous density $K$ to zero. As clearly seen in Eqs.~(\ref{eq:fcfulllast}), this comes from the coupling between the diagonal and off-diagonal matrix elements. For the fermionic case, the situation is different because the diagonal part of $K$ is automatically zero.
When we neglect only part of the components of $K$ for boson system, we obtain unphysical results. As an alternative to the previous approximation for boson system, one can also set
all components of $K$ to zero together with $M_{\alpha \beta}=0$ if $\alpha$ and $\beta$ are both in the environment. We compare in Fig.
\ref{fig:alldiagBBBF}  the approximate evolution with the full CEM evolution. By setting $K=0$, the unphysical evolution observed in Fig. \ref{fig:alldiag} for bosonic system disappears.
We see that such approximation is satisfactory reproducing the asymptotic behavior but some important physics is missed
during the evolution to equilibrium. Note that, this approximation can also be applied to the Fermi system (Fig. \ref{fig:alldiagBBBF}) but the reproduction of the full CEM approach degrades compared to the red lines in Fig. \ref{fig:alldiag}.

In the present section, we discussed the possibility to simplify the description of Fermi/Boson systems coupled to an environment by neglecting  the  components of the normal $M_{\alpha\beta}$ and anomalous $K_{\alpha\beta}$ densities when $\alpha$ and $\beta$ are both in the environment in accordance with the Eigenstate Thermalization Hypothesis (ETH)  \cite{Deu91,Sre94,Rig08,Rig09,San10,Rig12,Kha12}. It would be interesting to investigate further possible connection with the  ETH and its domain of different regime of validity
  when the quantum nature of the system change from Fermions to Bosons.

\subsection{Results for system coupled with two heat-baths }

Besides the possibility to treat mixture of fermions and bosons, one of the attractive aspects
of the present method  is the possibility to treat the coexistence of several baths. Equations (\ref{eq:fcfulllast}) to be solved
are unchanged when considering  several baths. The main difference is the sizes of the $M$ and $K$ matrices
that both increase with the number of baths after discretizing each environment.  We consider below a two-level system
coupled to two baths. Each bath, discretized using the method presented in section \ref{sec:dem} (see also \cite{Lac15}),
is characterized by the parameters $c_{i}$ and $\gamma_i$ as well as its initial temperature $T_i$ ($i=1,$ $2$).

\begin{figure}[!h]
\includegraphics[width= \linewidth]{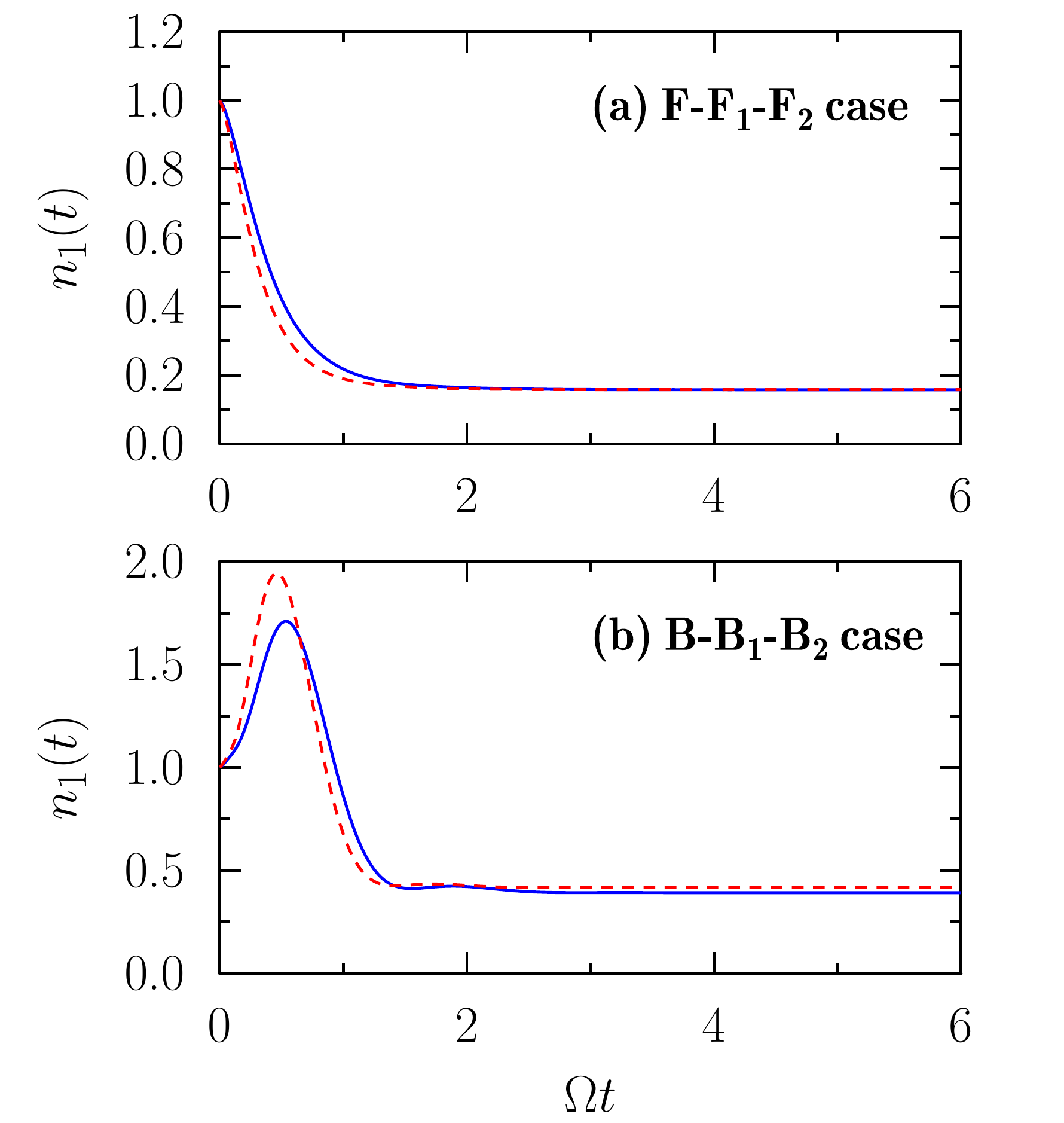}
    \caption{Evolution of $n_1(t)$ obtained in the FC case for the Fermi system coupled to two fermionic baths (a) (notation F-F$_1$-F$_2$) and
     for the Bose system coupled to two bosonic baths (b) (notation B-B$_1$-B$_2$) using the CEM approach. Two baths have the same temperature
     $T_1/\Omega=T_2/\Omega = 1$ and the same coupling strengths $c_1=c_2= 0.1$. The blue solid line
    corresponds to $\gamma_1/\Omega=\gamma_2/\Omega = 12$ while the red dashed line is obtained at $\gamma_1/\Omega=12$ and
    $\gamma_2/\Omega = 20$.
    }
    \label{fig:fffbbb}
\end{figure}

\subsubsection{System coupled to two heat baths of the same quantum nature}

As the first illustration, we consider the same conditions as  in  Ref. \cite{Hov18} where
the system is coupled to several heat-baths of the same quantum nature, i.e. the F-F$_1$-F$_2$ and
B-B$_1$-B$_2$ cases. This two cases are presented in Fig. \ref{fig:fffbbb} and can be compared to Fig.~3 of Ref. \cite{Hov18}.
For the B-B$_1$-B$_2$  case, we again perfectly recover our previous result. This could be considered as a numerical test
because the Laplace transform method used in Ref.~\cite{Hov18} and the present approach based on Eqs.~(\ref{eq:fcfulllast})
are strictly equivalent and exact.

For the case of fermionic system coupled to two fermionic baths, some differences are observed with \cite{Hov18}
although the global shape and asymptotic limit are similar.
In general, while the time-scale before reaching equilibrium and the asymptotic limit is globally in agreement
with Ref.~\cite{Hov18}, a difference is the absence of oscillations during the thermalization process.
We also checked numerically, as was analytically proved in our previous work, that the system coupled to two identical baths (the same $\gamma_i$ and the same temperature) can be treated as the system coupled to the single bath but with coupling strength equal to $c=c_1+c_2$.

\subsubsection{System coupled to two heat baths of mixed quantum natures}

\begin{figure}[!h]
\includegraphics[width= \linewidth]{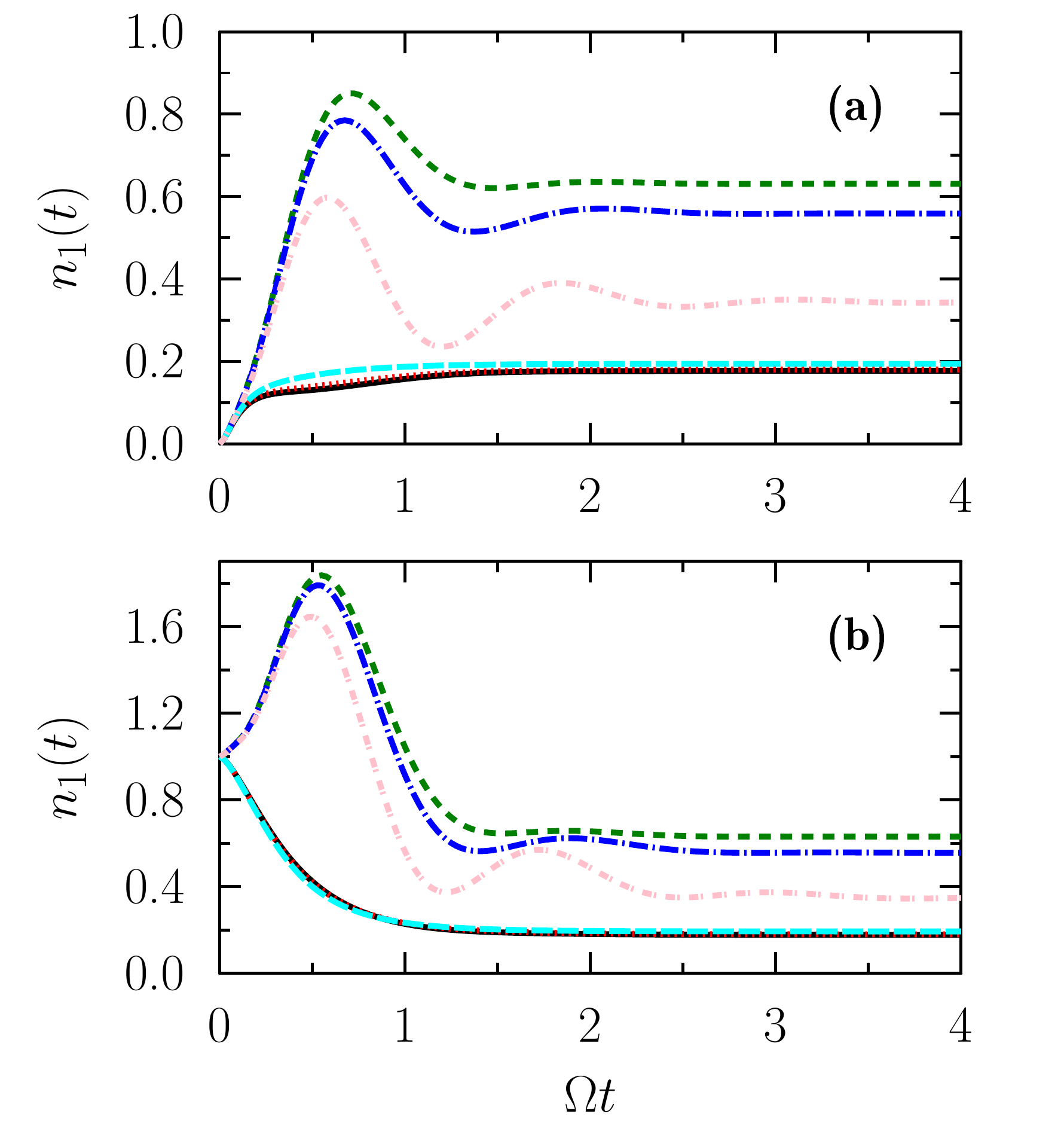}
    \caption{Evolution of $n_1(t)$ for the F-F$_1$-F$_2$ (black solid line), B-B$_1$-B$_2$ (green dashed line), F-B$_1$-F$_2$ (red dotted line), B-F$_1$-B$_2$ cases (blue dotted-dashed line), F-B$_1$-B$_2$ (cyan long-dashed line) and B-F$_1$-F$_2$ (pink short dashed-dotted line) obtained with the CEM approach and FC coupling starting from $n_1(0) = 0$ (a) and $n_1(0)=1$ (b).   The coupling strengths, spectral properties and temperatures are set to  $c_1=c_2=0.1$,  $\gamma_1/\Omega=10$, $\gamma_2/\Omega = 15$, $T_1/\Omega=1$, and $T_2/\Omega = 2.0$. In panel (b), the results for the F-B$_1$-F$_2$, F-B$_1$-B$_2$ and F-F$_1$-F$_2$ cases are almost identical.}
    \label{fig:mixed}
\end{figure}

We now analyze the change in the evolution compared to previous case when one of the bath or both baths have different quantum
natures compared to the system. Such situations are illustrated in Fig. \ref{fig:mixed}. To uncover the effect of changing Fermion into Boson or vice-versa, we compare situations where all other parameters
are unchanged, i.e. spectral function parameters, coupling strength and baths temperature remain  the same.

The first conclusion one could draw from Fig.~\ref{fig:mixed} is that, for the Fermi system, the nature of heat-bath does not affect much its evolution.
This is clearly the opposite to the bosonic system for which replacing bosonic bath by fermionic bath induces significant modification
both in the intermediate time evolution and asymptotic limit reached by the occupation probability.

Another aspect, which is visible in Fig.~\ref{fig:mixed}, is that an asymptotic stationary limit is always reached whatever are the natures of the system and heat-baths. This is in particular the case for the fermionic system coupled to two baths, one fermionic and one bosonic.
The convergence towards the stationary limit is systematically observed whatever are the properties of the baths, i.e. when changing the coupling strength, the spectral properties  and/or temperatures of two heat-baths. This conclusion is different from  the one we obtained previously for the F-B$_1$-F$_2$. Using slightly different approximation, we have shown that a stationary solution might never be reached \cite{Hov19}.  It should be noted however that both our previous prescription to the problem \cite{Sar17,Hov18,Sar18,Hov19,Hov20} and the
present one are only approximate when fermions are considered either in the system and/or baths.
The new approach proposed here has however the advantage
to account properly the Pauli principle for fermions. This could also be seen
from the occupation that remains bounded between 0 and 1. The possibility to reach or not a stationary limit when mixing baths with different
quantum natures is an interesting aspect. In particular, dedicated experiments would be interesting to clarify this issue.

 Because the bosonic systems coupled to one or several baths have been extensively investigated in our previous studies \cite{Sar14,Lac15,Sar17,Hov18,Sar18,Hov19,Hov20}, in the following, we focus on the Fermi systems and study in more details the evolutions
 obtained with the new approach proposed here.

 \subsection{Detailed study of Fermi system coupled to one or two bosonic and/or fermionic heat-baths}

 We systematically use Eqs.~(\ref{eq:fcfulllast}) to simulate the evolution of the Fermi two-level system coupled to
 one or two baths at various couplings, thermal and spectral properties. As illustrated in Fig.~\ref{fig:fffbbb} and \ref{fig:mixed},
 the evolution of $n_1(t)$ is rather simple and seems to correspond to the decay process.
 Based on this observation and with the goal to infer generic
 properties of the system evolution due to the surrounding environment, we fit the occupation number evolution $n_f(t)=n_1(t)$
 with the simple function
\begin{eqnarray}
n_f(t) = n_f(\infty) + [n_f(0) - n_f(\infty) ] e^{-\Gamma_f t}.  \label{eq:nffit}
\end{eqnarray}
In the following study, we will consider the case $n_f(0) = 1$. $n_f(\infty) $ and $\Gamma_f$ are fitted on the evolutions
and correspond respectively to the asymptotic occupation number and to the
decay time $\tau_f = 1/\Gamma_f$. Despite its simplicity, Eq.~(\ref{eq:nffit}) turns out to provide a rather precise description of the evolution for the whole range of couplings and temperatures considered.
\begin{figure}[!h]
\includegraphics[width= \linewidth]{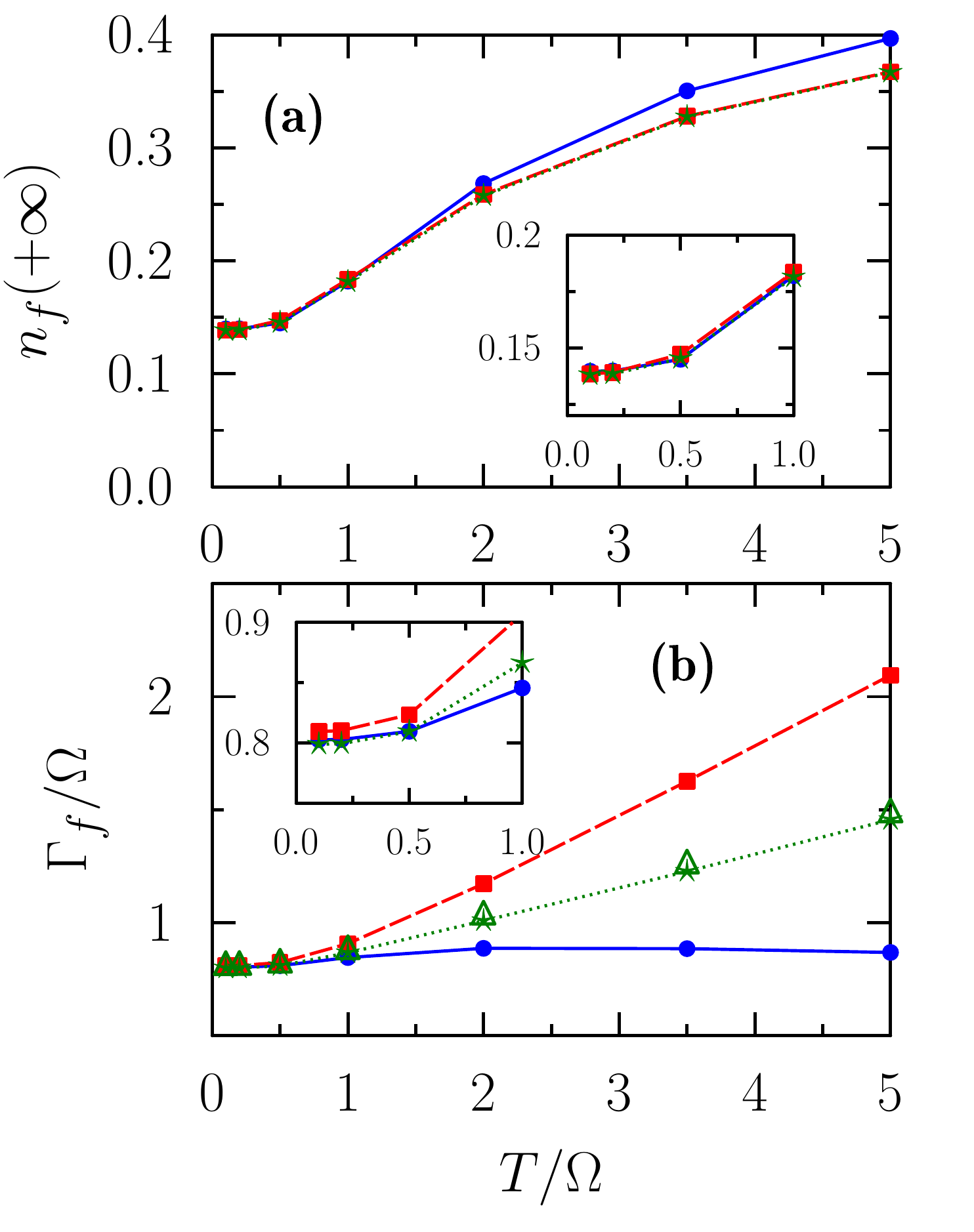}
    \caption{Evolution of $n_f(\infty) $ (a) and $\Gamma_f$ (b) obtained by fitting the evolution of $n_f(t)$ with expression (\ref{eq:nffit}) for the
     Fermi system coupled to one or two baths with various initial temperatures.
    The blue circles correspond to the F-F case, where the bath properties are set to $c_1=0.1$ and
    $\gamma_1/\Omega=12$. The red squares correspond to the F-B case where the only difference with previous case is that the
    Fermi bath is replaced by the Bose bath.  The green stars correspond to the F-B$_1$-F$_2$ case with $c_1=c_2=0.05$,
    and $\gamma_1/\Omega=\gamma_2/\Omega=12$. Note the F-B$_1$-B$_2$ and F-F$_1$-F$_2$ (not shown) match exactly the F-B and F-F case with a coupling equal to $c_1+c_2$. In the multi-baths case, we assume that all baths are at the same temperature, $T_1=T_2=T$. In both panels, the inset is a focus on the low temperature limit. The green open squares corresponds
    to the result of Eq. (\ref{eq:linc}). }
    \label{fig:FBFT}
\end{figure}
The systematic evolutions of $\Gamma_f$ and $n_f(\infty)$ obtained in the presence of one or two baths are reported
in Fig.~\ref{fig:FBFT}.  

\begin{figure}[!h]
\includegraphics[width= \linewidth]{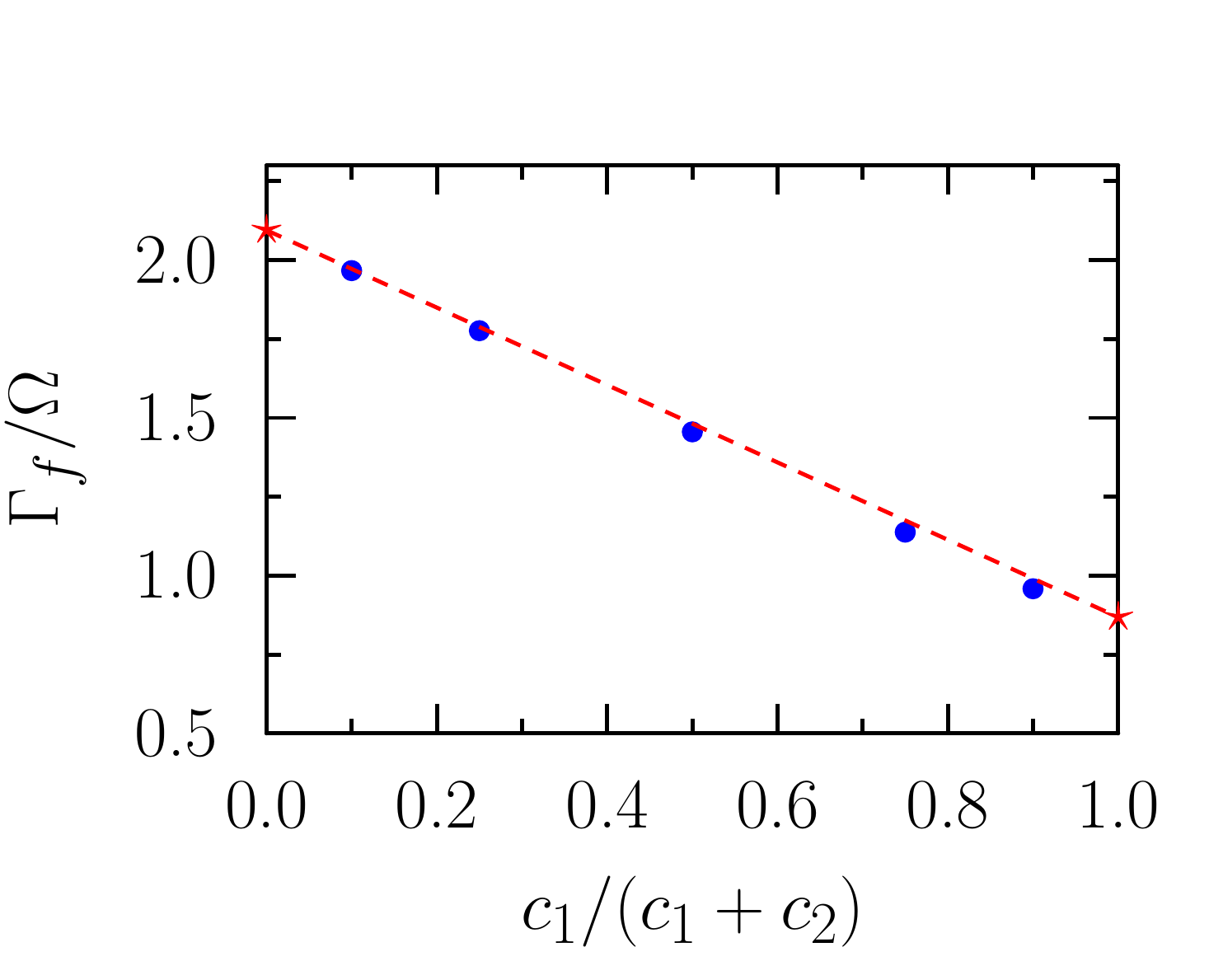}
    \caption{Evolution of $\Gamma_f$ obtained for the F-B$_1$-F$_2$ as a function of $c_1/(c_1+c_2)$ assuming that $c_1+c_2=0.1$. The structural properties of the two baths are the same as in Fig. \ref{fig:FBFT} while $T_1/\Omega = T_2/\Omega=5$. The stars
    at the extremes correspond to the reference F-F  (F-B) calculations put artificially at $c_1/(c_1+c_2) = 1$ (0). The dashed line is a linear interpolation between the two stars. }
    \label{fig:g1g2}
\end{figure}

Focusing first on the single bath case, we  observe that the F-F and F-B cases lead to rather similar properties at low temperature. The evolutions of the Fermi systems at low temperature ($T/\Omega < 1$) appear to be rather insensitive to the quantum nature of the bath and/or if one- or several baths are coupled to the system. When the
temperature of the bath(s) increases, we see  significant differences in $\Gamma_f$ and to a lesser extent in $n_f(\infty)$ depending on the quantum nature of the baths.
For $n_f(\infty)$, we see only small differences between the F-F and F-B cases. It is interesting however to mention that the asymptotic occupation number of the F-B$_1$-F$_2$ matches the one of the F-B case for all temperatures. Therefore,  for two baths with different quantum natures but with equivalent spectral properties, the
bosonic bath seems to decide the asymptotic behavior.

In the F-F case, the decay time is almost independent of the temperature while for the F-B case, $\Gamma_f$  linearly increases with temperature
at $T/\Omega > 1$. This implies that the transient time to equilibrium is much shorter if the fermionic system is coupled to  bosonic bath rather than to  fermionic bath.
We also observe in Fig.~\ref{fig:FBFT} that $\Gamma_f$ obtained for the F-B$_1$-F$_2$ case is in-between the F-F and F-B cases. More precisely, we have $\Gamma_f = (\Gamma_{\rm FF} + \Gamma_{\rm FB})/2$ where we use the notations
$\Gamma_{\rm FF}$ ($\Gamma_{\rm FB}$) for the value of $\Gamma_f$ obtained in the F-F (F-B) case.   We further
investigated this simple behavior in the  F-B$_1$-F$_2$  case by varying the couplings values of $c_1$ and $c_2$ while keeping $c_1+c_2=0.1$.
The evolution of
  $\Gamma_f$  is displayed as a function of $c_1/(c_1+c_2)$ in Fig.~\ref{fig:g1g2} at $T_1/\Omega=T_2/\Omega=5$.
  We observe in this figure that we have approximately:
 \begin{eqnarray}
\Gamma_f &\simeq& \frac{c_1}{(c_1+c_2)} \Gamma_{\rm FF} +  \frac{c_2}{(c_1+c_2)} \Gamma_{\rm FB},  \label{eq:linc}
\end{eqnarray}
that is a very simple relationship.

Note that if we change the temperature, we change the absolute value of the dependence presented in Fig.~\ref{fig:g1g2}, but  all conclusions remain valid.
Such a behavior can again be tested in the experimental observations. In particular, one might imagine by mixing several baths and changing the relative strengths of the couplings with the system to control the decay properties even if the temperatures of different baths
are kept fixed.

The approximate treatment of an open quantum Fermi system discussed in section  \ref{sec:appfermi} and Appendix \ref{sec:approx} turns out to be useful to understand qualitatively
the conclusion made from Figs.~\ref{fig:FBFT} and \ref{fig:g1g2}. We have checked numerically that the approximation is also
accurate when the Fermi system is coupled to several baths in the high temperature--weak coupling limit.
As shown in Appendix \ref{sec:approx}, the simple linear relation (\ref{eq:linc}) can be explained consistently with this simplification.
Indeed, starting from the analytical equations (\ref{eq:gamd}) for  the asymptotic occupation probability and decay time,
one can explain why the decay time is independent of temperature in the case of Fermi bath while it increases with the temperature for Bose bath.




\section{Conclusion}

We proposed here a new approach called CEM to describe the system coupled to one or several baths eventually mixing different quantum natures
of the particles (fermions or bosons). The approach is exact when only bosonic degrees of freedoms are considered and
provide an approximate solution when fermions are also present either in the system and/or in the bath. In this novel approach,
a particular attention is paid to properly account for the Pauli principle for the fermions.
The approach is illustrated for the system coupled to single bath where the system or bath can be either fermionic or bosonic.
For Fermi systems, we showed that the proper treatment of the Fermi nature is essential to obtain an accurate
treatment of the evolution.

The approach includes non-Markovian effects and is rather versatile and we do not anticipate any specific difficulty to apply it to the baths with complex structure. We  illustrated the method for a system coupled to two baths with various quantum statistics. One of the important aspects that differs from our previous solution \cite{Hov19,Hov20} to this problem is that Fermi systems coupled to two baths always reach an asymptotic stationary limit. Since in the present
work, similarly to Ref.~\cite{Hov19}, the approach we propose is not exact, the existence or not of an asymptotic time-independent solution
is an interesting debate. In particular, it would be interesting to give an experimental clarification to this aspect.
Besides the asymptotic behavior, we observe that Fermi systems have relatively simple decay properties compared to Boson systems coupled to the same baths. We showed that the decay time of the Fermi system coupled to the fermionic and bosonic baths can be easily related to the cases of the system coupled to only one bosonic bath or to only one fermionic bath.

In the present work, we focused our attention to a single Qubits  coupled to a set of environments including fully non-Markovian effects.
The theory can a priori be extended to obtained numerical simulation of an ensemble of Qubits with the price of increasing the numerical cost.
With the target goal to be able to treat eventually several hundreds  of Qubits we also explore the possibility to obtain simplified theories
of Fermi systems while not degrading the description of evolution. We show that in some regime of coupling or temperature, the
simplification can indeed be made.

\section*{Acknowledgments}
The IN2P3(France)-JINR(Dubna) Cooperation Programme is gratefully acknowledged.
This work was partially supported by  Russian Foundation for Basic Research (Moscow), N 17-52-12015 and 20-02-00176.
This project has received financial support from the CNRS through the 80Prime  program.

\appendix

\begin{widetext}

\section{Simplified form of equations of motion for Fermi system coupled to one or several baths}
\label{sec:approx}

We consider here the case of  Fermi system coupled to one or several baths.  Assuming $K_{\alpha\beta} (t) = M_{\alpha\beta} (t) = 0$
when $\alpha$ and $\beta$ components are in the bath  with $\alpha \neq \beta$
and introducing the notations:
\begin{eqnarray}
n_\alpha(t) &=& M_{\alpha\alpha}(t), ~~~ \bar n_\alpha (t) = 1 + \varepsilon_\alpha n_\alpha(t), 
\end{eqnarray}
Eqs.~(\ref{eq:fcfulllast}) are simplified as:
  \begin{eqnarray}
\left\{
\begin{array}{ll}
\displaystyle
\dot n_1(t) &=  i \sum_\nu g_\nu (M_{\nu 1}  - M_{1\nu}) +  i \sum_\nu g_\nu (K^*_{1\nu} -
K_{1\nu} )  \\
\\
\dot n_\alpha(t) &=  i g_\alpha (M_{1\alpha} - M_{\alpha 1}) +ig_\alpha (K^*_{1\alpha} - K_{1\alpha})\\
\\
\displaystyle \dot M_{1\alpha} &=      i  (\omega_1 - \omega_\alpha) M_{1\alpha}
+i g_\alpha \left[ \bar n_1(t) n_\alpha (t)  -  n_1(t) \bar n_\alpha (t)\right]   \\
\\
\displaystyle \dot K_{1\alpha} &= i  (\omega_1 + \omega_\alpha)  K_{1\alpha}
+i g_\alpha \left[ \bar n_1(t) \bar n_\alpha (t)  -  n_1(t) n_\alpha (t)\right]
\end{array}
\right.  .
\label{eq:simp10}
\end{eqnarray}
The numerical integration of these equations of motion is much less demanding than the original  set of equations (\ref{eq:fcfulllast}). This could be seen from the fact that the original number of coupled equations was $2N^2_{\rm tot}$ where
$N_{\rm tot}$ is the total number of creation/annihilation operators for the system+baths, while the number of coupled equations in (\ref{eq:simp10}) is reduced to  $(3 N_{\rm tot} -2)$.

\subsection{Pauli Master equation with memory effect}
Related approximation has been discussed for instance in Ref. \cite{Sar18,Hov18}. Following these references,  one might eventually obtain
the closed form of the equations for $n_1(t)$ and  $n_\alpha(t)$ by formally integrating the last two equations.  Using the fact that
$M_{1\alpha}(0) = K_{1\alpha} (0)= 0$, we have the formal solution:
\begin{eqnarray}
i g_\alpha [ M_{\alpha 1} (t) - M_{1 \alpha}(t) ]&=& 2 g^2_\alpha  \int_0^t  \cos([\omega_1 - \omega_\alpha] [t-\tau])\left[ \bar n_1(\tau) n_\alpha (\tau)  -  n_1(\tau) \bar n_\alpha (\tau)\right], \nonumber \\
i g_\alpha [ K^*_{1\alpha}(t) - K_{1\alpha}(t)] &=& 2 g^2_\alpha  \int_0^t  \cos([\omega_1 + \omega_\alpha] [t-\tau])\left[ \bar n_1(\tau) \bar n_\alpha (\tau)  -  n_1(\tau) n_\alpha (\tau)\right]. \nonumber
\end{eqnarray}
The evolution of the system occupation probabilities is written as the Pauli master equation:
 \begin{eqnarray}
\frac{d n_1(t)}{dt} &=& \int_0^t \left\{  {\cal W}_+^1(t,\tau)   \bar n_1(\tau) - {\cal W}_-^1(t,\tau)   n_1(\tau) \right\} d\tau,  \label{eq:dn1}
\end{eqnarray}
with
\begin{eqnarray}
 {\cal W}_+^1(t,\tau)  &=& 2 \sum_\alpha g_\alpha^2
 \left[ \cos([\omega_1 - \omega_\alpha] [t-\tau]) n_\alpha (\tau) + \cos([\omega_1 + \omega_\alpha] [t-\tau]) \bar n_\alpha(\tau) \right] , \nonumber \\
  {\cal W}_-^1(t,\tau)  &=& 2 \sum_\alpha g_\alpha^2
 \left[ \cos([\omega_1 - \omega_\alpha] [t-\tau]) \bar n_\alpha (\tau) + \cos([\omega_1 + \omega_\alpha] [t-\tau])  n_\alpha(\tau) \right] \nonumber . 
\end{eqnarray}
These expressions can eventually be complemented by the set of equivalent master equations for the $n_\alpha (t)$ (not shown here).
In the following, we use the notations $G_1(\tau) =  {\cal W}_+^1(t,\tau)  + {\cal W}_-^1(t,\tau) $   and
$F_1(\tau) =   {\cal W}_+^1(t,\tau) - {\cal W}_-^1(t,\tau)$. Some simple manipulations result in the expressions:
\begin{eqnarray}
G_1(t,\tau) &=& 4 \cos(\omega_1 \tau) \sum_\alpha g_\alpha^2  \cos(\omega_\alpha \tau) [n_\alpha(\tau) + \bar n_\alpha(\tau)] ,  \label{eq:g1}\\
F_1(t,\tau) & = & 4 \sin(\omega_1 \tau) \sum_\alpha g_\alpha^2  \sin(\omega_\alpha \tau) [n_\alpha(\tau) - \bar n_\alpha(\tau)]. \label{eq:f1}
\end{eqnarray}

\subsection{Simple approximate form for system evolution in weak-coupling regime}

Starting from the master equation (A2), one might try to see if the equation of motion of the system occupation number
can be written in terms of a time-local equation while properly keeping non-markovian effects. Our goal is to make
connection with the simple form (\ref{eq:nffit}) used to fit the $n_1(t)$ evolution.
For this, we first rewrite the master equation  as
 \begin{eqnarray}
\frac{d n_1(t)}{dt} &=& - \int_0^t G_1(t,\tau) n_1(\tau) d\tau
+  \int_0^t   {\cal W}_+^1(t,\tau) d\tau , \label{eq:dn1bis}
\label{eq:dnalpha}
\end{eqnarray}
where we use the fact that the system is fermionic. Note that $G_1(t,\tau)$ is linked to the decay time of the Fermi system.
If the bath contains only fermions, i.e. for all $\alpha$ we have $n_\alpha(\tau) + \bar n_\alpha(\tau)=1$, this quantity
becomes independent of the initial temperature of the baths. Starting from this expression,
using the expression for $g_\alpha$ given in section \ref{sec:dem}, and taking the continuous limit for the bath, we deduce for a
fermionic bath
\begin{eqnarray}
G_1(t,\tau) &=& 4 \cos(\omega_1 \tau) \int_0^{+\infty} d\omega J_1 (\omega)  \cos(\omega \tau)
 \end{eqnarray}
 where $J_1(\omega)$ is the spectral function (\ref{eq:lorentz}).

From now on, we assume that the coupling between the system and bath is weak enough and we can only retain terms up to the second order
in $g^2_\alpha$. Consistently with this approximation, one might eventually make the replacement $n_1(\tau) \simeq n_1(t)$ together
with $n_\alpha(\tau) \simeq n_\alpha(0)$ in the integral in time such that we obtain
\begin{eqnarray}
\frac{d n_1(t)}{dt} \simeq - \Gamma_f(t) n_1(t) + D_f (t),
\end{eqnarray}
where we use
\begin{eqnarray}
\Gamma_f(t) &\simeq &    \int_0^t \overline{G_1}(\tau) d\tau, ~~~
D_f(t) \simeq   \frac{1}{2} \int_0^t \left[ \overline{G_1}(\tau) + \overline{F_1}(\tau) \right] d\tau , \label{eq:gamd}
\end{eqnarray}
with
\begin{eqnarray}
 \overline{G_1}(\tau) &=& 4 \cos(\omega_1 \tau) \sum_\alpha g_\alpha^2  \cos(\omega_\alpha \tau) [n_\alpha(0) + \bar n_\alpha(0)] ,  \nonumber\\
 \overline{F_1}(\tau) & = & 4 \sin(\omega_1 \tau) \sum_\alpha g_\alpha^2  \sin(\omega_\alpha \tau) [n_\alpha(0) - \bar n_\alpha(0)]. \nonumber
\end{eqnarray}
Equations (A8) and (A9) can be generalized to the case of several baths.

From these expressions, we see that $ \overline{G_1}(\tau)$  and therefore $\Gamma_f$ calculated with Eqs.~(\ref{eq:gamd})
are independent of temperature for fermion bath (Fig.~\ref{fig:FBFT}). However, $\Gamma_f$ depends
on temperature for bosonic bath. 

Note finally that, in the simplified scenario presented here one deduces a simple time-local equation, valid a priori in the weak-coupling regime.
This equation however includes partially non-Markovian effects. As a side remark, it would be interesting to investigate the possibility to extend the time-convolutionless approach of Refs.~\cite{Has77,Shi77,Bre01,Gem07, Bre02} to obtain a systematic constructive framework leading  to the time-local equation of motion for the system with higher orders corrections in the coupling.

\end{widetext}

\end{document}